\documentclass[usletter,12pt,final,titlepage]{article}

\usepackage{amsfonts,amsmath,amsthm,amssymb,xcolor,colortbl}
\usepackage{enumitem}
\usepackage{calrsfs}
\DeclareMathAlphabet{\pazocal}{OMS}{zplm}{m}{n}

\usepackage{geometry}
\usepackage[onehalfspacing]{setspace}
\usepackage{soul}
\usepackage{setspace}
\usepackage{natbib}
\usepackage{clipboard}	%
\usepackage{graphicx}
\newclipboard{PaperClipboard}	%
\usepackage{xr-hyper} %
\usepackage[colorlinks,linkcolor=links,citecolor=cites,urlcolor=MyDarkBlue]{hyperref}
\usepackage{cleveref}
\usepackage{bbm}
\usepackage[justification=centering]{caption}
\usepackage[justification=centering]{subfig}
\usepackage[section]{placeins}
\usepackage{parskip}
\usepackage{nicefrac}
\usepackage{appendix}
\usepackage[multiple]{footmisc}
\definecolor{MyDarkBlue}{rgb}{0,0.08,0.45}
\definecolor{cites}{HTML}{324b13}
\definecolor{links}{HTML}{1a663b}
\definecolor{MyLightMagenta}{cmyk}{0.1,0.8,0,0.1}
\definecolor{sblue}{HTML}{0049A9}
\definecolor{scyan}{HTML}{CBEAFC}
\definecolor{sred}{HTML}{B5595C}
\definecolor{sgreen}{HTML}{609B57}
\definecolor{spink}{HTML}{FFB0FF}

\DeclareMathOperator*{\argmin}{arg\,min}
\usepackage{mparhack}

 \usepackage{chngcntr}
\usepackage{apptools}
\AtAppendix{\counterwithin{equation}{section}}
\AtAppendix{\counterwithin{figure}{section}}
\AtAppendix{\counterwithin{table}{section}}
\AtAppendix{\counterwithin{example}{section}}
\AtAppendix{\counterwithin{remark}{section}}
\AtAppendix{\counterwithin{proposition}{section}}
\AtAppendix{\counterwithin{theorem}{section}}
\AtAppendix{\counterwithin{corollary}{section}}
\usepackage{etoolbox}
\makeatletter
\patchcmd{\hyper@makecurrent}{%
    \ifx\Hy@param\Hy@chapterstring
        \let\Hy@param\Hy@chapapp
    \fi
}{%
    \iftoggle{inappendix}{%true-branch
        \@checkappendixparam{chapter}%
        \@checkappendixparam{section}%
        \@checkappendixparam{subsection}%
        \@checkappendixparam{subsubsection}%
        \@checkappendixparam{paragraph}%
        \@checkappendixparam{subparagraph}%
    }{}%
}{}{\errmessage{failed to patch}}

\newcommand*{\@checkappendixparam}[1]{%
    \def\@checkappendixparamtmp{#1}%
    \ifx\Hy@param\@checkappendixparamtmp
        \let\Hy@param\Hy@appendixstring
    \fi
}
\makeatletter

\newtoggle{inappendix}
\togglefalse{inappendix}

\apptocmd{\appendix}{\toggletrue{inappendix}}{}{\errmessage{failed to patch}}
\apptocmd{\subappendices}{\toggletrue{inappendix}}{}{\errmessage{failed to patch}}

\newcommand{\eps}{\varepsilon}

\newcommand{\E}{\mathbb{E}}
\newcommand{\Prob}{\mathbb{P}}

\def\OL{The OPO}

\counterwithin{figure}{section}
\counterwithin{equation}{section}

\title{Social Learning in Lung Transplant Decision\thanks{The data reported here have been supplied by the Hennepin Healthcare Research Institute (HHRI) as the contractor for the Scientific Registry of Transplant Recipients (SRTR). The interpretation and reporting of these data are the responsibility of the author(s) and in no way should be seen as an official policy of or interpretation by the SRTR or the U.S. Government.}}

\author{Laura Doval\thanks{Columbia Business School and CEPR. E-mail: \href{mailto:laura.doval@columbia.edu}{\texttt{laura.doval@columbia.edu}}} \and Federico Echenique\thanks{University of California, Berkeley. E-mail: \href{mailto:fede@econ.berkeley.edu}{\texttt{fede@econ.berkeley.edu}}} \and Wanying Huang\thanks{Department of Economics, Monash University. E-mail: \href{mailto:kate.wanyinghuang@gmail.com}{\texttt{kate.huang@monash.edu}}} \and Yi Xin\thanks{California Institute of Technology. E-mail: \href{mailto:yixin@caltech.edu}{\texttt{yixin@caltech.edu}}}
}

\date{\today}

\begin{document}

\maketitle

\begin{abstract}
We study the allocation of deceased-donor lungs to patients in need of a transplant. Patients make sequential decisions in an order dictated by a priority policy. Using data from a prominent Organ Procurement Organization in the United States, we provide reduced-form evidence of social learning: because patients accept or reject organs in sequence, their decisions exhibit herding behavior, often rejecting an organ that would otherwise be accepted. We develop and estimate a structural model to quantify the impact of various policy proposals and informational regimes. Our results show that blinding patients to their position in the sequence\textemdash thereby eliminating social learning\textemdash boosts organ allocation but reduces average utility per patient. In contrast, prioritizing patients by their likelihood of acceptance exacerbates social learning, leading to fewer organ allocations. Nevertheless, it raises utility per accepted organ and expedites the allocation process.


\end{abstract}

\newpage

\section{Introduction}

We study the allocation of deceased-donor lungs to patients in need of a transplant. An organ is offered sequentially to a list of patients according to a priority order.  Within a small window of time, each patient, in consultation with their clinical coordinators and medical doctors, needs to assess both the organ's quality and their compatibility with it to decide whether to accept the offer.\footnote{A patient's medical team usually makes the final decision in consultation with the patient. As is standard in the literature, we assume the medical team has the best interests of the patient in mind when making this decision. For simplicity, we refer to this collective decision-making unit as the \emph{patient}.} Deciding whether to accept an organ offer is one of the most important decisions made by a patient, and one that is marred with uncertainty.  Rejecting a high-priority offer increases the likelihood that the patient will leave the system unmatched, prolonging their compromised quality of life. Conversely, accepting a low-priority offer with potentially poor match quality can reduce the patient's chances of survival post-transplant \citep{gilroy20192016}.

To facilitate decision-making, patients are provided with information about donor characteristics, often including their own position in the order of offers for the organ. In particular, when a lung is offered to a patient, the patient is informed of their \emph{sequence number in the match run}\textemdash which indicates their priority, with smaller numbers representing higher priority\textemdash as well as the reasons for any previous offer rejections \citep{harhay2019donor}. Simply put, patients with a smaller sequence number receive an offer before patients with a larger sequence number. Recently, the United Network for Organ Sharing (UNOS) has ensured that information about the sequence number of the patient is readily and systematically available for all types of organs, including lungs.\footnote{See \href{https://unos.org/news/match-summary-view-now-available-in-donornet-desktop/}{https://unos.org/news/match-summary-view-now-available-in-donornet-desktop/}.}

At the same time, rejections early in the match run often lead to further rejections later in the sequence \citep{gilroy20192016}, motivating policy proposals to hide match run positions from patients \citep{cohen2018position,benvenuto2019should}. This is because, when an organ is rejected early in the match run, patients later in the sequence may infer that the organ is of lower quality. %
Since organ quality is the most frequently cited reason for rejecting an organ offer, receiving an offer at a lower priority may indicate a harder-to-place organ \citep{goldberg2016liver}. Therefore, while providing information about early rejections could facilitate decision-making, it may also be detrimental to organ placement if this information crowds out the patient's independent assessment of the organ.

In this paper, we use data from a large organ procurement organization to assess how the availability of information on sequence numbers affects organ acceptance rates, and patient welfare, in the allocation of deceased-donor lungs. We make two contributions. First, we provide reduced-form evidence of \emph{social learning}: %
having a lower priority in the match run alone\textemdash regardless of the factors that determine this priority\textemdash makes patients more likely to reject the organ compared to those with higher priority. This is consistent with ``herding,'' the main predictions of social learning models. Past rejections by higher-priority patients lead lower-priority patients to infer that the organ is of low quality, irrespective of their own assessment of the organ. In consequence, a few rejections early in the match run trigger (with high probability) long spells of inefficient rejections. Organs can go to waste even when there are patients who would find them acceptable. Second, we estimate a structural model of sequential assignment, and use the model estimates to evaluate the welfare implications of different priority and information policies in the presence of social learning.

Evidence of social learning is presented in \autoref{sec:reduced_form}. We present reduced-form evidence that information about match run position negatively affects the probability of organ placement. After controlling for all donor and patient risk factors (including those measuring donor-patient compatibility), we find that, conditional on patients who initially express interest in accepting the organ, their probability of final acceptance decreases as the sequence number increases. These results suggest that organs offered at larger sequence numbers are perceived as lower quality. Meanwhile, our data reveals a significant fraction of unused organs: 47\% of lungs are ultimately rejected by all matched patients, and thus remain unallocated.\footnote{Unlike the allocation of deceased-donor kidneys, where organs are routinely recovered before being allocated and accepted, lungs that are not accepted are not recovered and remain in the donor.} Since a match run is only initiated for organs that are deemed viable, this high rate of unused donor organs suggests significant potential for improving efficiency. %

Motivated by our reduced-form findings, in \autoref{sec:model} we propose a structural model of sequential assignment. Our model adapts the canonical social learning model \citep{banerjee1992simple, BichHirshWelch:92} to the institutional details of deceased-donor organ allocations. Specifically, when matched with a donor, each patient is assigned a sequence number that determines the order in which they decide whether to accept or reject the offer. In addition to the observed donor characteristics, there is a common unknown component of organ quality that can be either \emph{high} or \emph{low}. Before making their decision, each patient receives a private, imperfect binary signal about the organ quality,\footnote{Recall that we use ‘patient’ to refer collectively to the patient and their medical team, so this signal could reflect information available to the medical team.} and observes all previous rejections by patients with lower sequence numbers. They then decide to accept if, based on all available information, their expected utility from accepting is higher than from rejecting.
In \autoref{sec:estimation}, we present our model estimates, which reveal a high prior probability of low-quality organs and a high precision in patient's private signals. Our estimated model closely replicates both the data and the reduced-form evidence.

Using our estimated model, we conduct counterfactual experiments to assess the welfare implications of different priority policies, and informational regimes, while accounting for social learning. The first set of counterfactual experiments examines two contrasting priority ranking policies: under the \emph{greedy priority} policy, patients are ranked in decreasing order of their probability of accepting the organ in question, whereas under the \emph{reverse greedy priority} policy, patients are ranked in increasing order. Since the greedy policy offers organs to patients who value them the most (and thus have the highest acceptance probability), it may seem at first to lead to faster acceptances. Social learning, however, complicates this dynamic: a rejection by a patient with higher utility for the organ conveys a stronger negative signal about the organ's quality than if the same decision was made by a lower-utility patient. This negative effect of social learning is amplified under the greedy priority policy, as a series of rejections from higher-utility patients reinforces doubts about the organ, which slows down further acceptances. Thus, a priori, it is unclear whether the greedy priority policy, when interacting with social learning, leads to a more efficient allocation process. 

\autoref{tab:counter_efficiency} summarizes the results from our first set of counterfactual experiments. It shows that the greedy priority policy results in faster allocations as organs are, on average, accepted by patients with smaller sequence numbers. However, due to the negative informational effect of social learning, this policy also leads to fewer organs being allocated compared to the current policy. Nevertheless, patients who ultimately receive organs under this policy are better off on average, as they value the organ highly and are less likely to receive low-quality organs. Thus, while the greedy policy results in a slightly lower allocation rate because of social learning, it ensures that organs are allocated quickly and to patients who value them more. In contrast, the reverse greedy policy allocates more organs than the current policy, but at a slower rate. In addition, the average utility of patients who receive a transplant is lower under reverse greedy. This reduction
occurs partly because early rejections under the reverse greedy policy provide little information about organ quality, which then leads to a higher acceptance rate of low-quality organs. Of course, a low-quality organ may still be acceptable to many patients, since in principle all organs that are offered in a match run are deemed medically viable.

\autoref{tab:counter_efficiency_info} presents the results from our second set of counterfactual experiments, which introduce different informational regimes under the current priority policy. First, we implement a \emph{no social learning} regime, where patients are blinded to their position in the match run when making decisions.\footnote{If there are multiple acceptances, the organ is allocated to the patient with the smallest sequence number among those who accept the offer.} Since each patient decides based solely on their own private signal, this regime eliminates the negative effects of social learning. As a result, it achieves the highest acceptance rate by uniformly increasing the acceptance of all organs, regardless of their quality. However, due to the lack of information aggregation, it also yields the lowest utility among patients who receive a transplant and results in the slowest allocation process. Second, we consider an \emph{information sharing} regime, where each patient has access to the private signals of all higher-priority patients in the same match run. By making past private signals publicly available, this regime allows for the most amount of information aggregation (not just negative information) compared to both regimes with and without social learning, serving as a useful benchmark. Consequently, it achieves the highest average utility for patients who receive a transplant and the fastest allocation process. 

Altogether, our results highlight the trade-offs involved in providing patients with information about their match run position and how this interacts with different priority assignment policies. Our evidence shows that rejections in a match run convey information about organ quality beyond the factors used to determine a patient's priority. Specifically, knowing their match run positions allows patients to infer additional information from early rejections; however, these early rejections can distort the underlying information, and, in the presence of social learning, lead to excessive rejections. Thus, this finding supports proposals for organ procurement organizations to share more comprehensive and detailed information with patients and their medical teams. The more information available to the patients, the less they will need to rely on inferences based on the limited information obtained from others' decisions. Taking into account the effect of social learning, our counterfactual analysis suggests that the greedy priority policy performs the best in terms of maximizing patient utility per acceptance and expediting the allocation process.

\paragraph{Related Literature} We are closest to the contribution of \cite{zhang2010sound}, who studies the role of social learning in the allocation of deceased donor kidneys. Our result that, all else equal, the patient's sequence number negatively affects the organ acceptance probability mirrors a similar observation in \cite{zhang2010sound}. Through counterfactual simulations, the author shows that eliminating social learning reduces organ wastage. More recently, \cite{de2020herding} study how transplant centers assess organ quality and how variability in their assessment abilities impacts the matching process between donors and transplant centers. In contrast to \cite{zhang2010sound}, \cite{de2020herding} find that eliminating the channel of social learning across transplant centers does not significantly improve the organ discard rate.\footnote{Social learning has been empirically studied in other contexts (see, e.g., \cite{cipriani2014estimating} for a financial application, and \cite{newham2018herd} for a voting application within the FDA).} This paper complements previous works by focusing on the allocation of deceased-donor lungs. We find that while the accumulation of negative information through social learning decreases the organ acceptance rate, it also facilitates the rejection of low-quality organs. In the absence of social learning, there is less information aggregation, which results in lower patient utility and a slower allocation process. Thus, there is a trade-off between maximizing the allocation rate and achieving a system that is both efficient and high in match quality.

Several market design papers have empirically studied the efficiency and welfare properties of deceased donor organ allocation, proposing various improvements. For instance, focusing on the allocation of deceased-donor kidneys, \cite{agarwal2018dynamic,agarwal2021equilibrium} study how dynamic incentives, where forward-looking patients strategically wait longer to secure more desirable organs, affect organ wastage. \cite{shi2022eliminating} show that eliminating organ wastage is at odds with prioritizing patients on the basis of their waiting time. \cite{munoz2021incentive} studies the effects of asymmetric information in the allocation of hearts and livers, in which doctors are better informed about patients' urgency and readiness status than the organ procurement agency. 

Lastly, the connection between match run position at which an organ is accepted and transplantation outcomes, such as mortality and graft survival, is extensively studied in the medical literature (see, e.g., \citealp{cohen2018position} for kidneys, \citealp{goldberg2016liver} for liver, and \citealp{harhay2019donor} for lungs). Most of these studies conclude that while there is a negative correlation between 
transplantation outcomes and match run position at which an organ is accepted, these correlations are not statistically significant once one controls for the transplant center that accepts the organs. The conclusion is then that transplant centers that accept organs at later positions, which are typically higher volume centers, have more experience in assessing organ quality and are willing to take more organs because they have a larger patient population. 

This strand of medical literature, however, acknowledges several limitations: (i) the analysis is conducted solely on organs that are eventually transplanted, without accounting for selection bias; (ii) other important outcomes, such as hospital length of stay, duration of mechanical ventilation, postoperative functional status, quality of life, and transplant cost, are not considered; and (iii) the conclusions rely on the assumption that the data about donor-risk factors available to the econometrician is also available to the transplant center at the time of offer evaluation\textemdash an assumption that conflicts with many recent policy proposals aimed at making donor data more readily accessible to evaluating transplant centers \citep{gilroy20192016,hackmann2022improving}. Due to these limitations, the medical community has requested a randomized control trial in which some transplant centers are blinded to their patient position in the match run to meaningfully evaluate the impact of match run position on transplant outcomes \citep{cohen2018kidney}.

\paragraph{Organization} The rest of the paper is organized as follows. \autoref{sec:inst-data} describes the institutional setting and our data. \autoref{sec:reduced_form} provides reduced form evidence of social learning in the allocation of deceased donor lungs. \autoref{sec:model} presents the structural model, and \autoref{sec:estimation} discusses the results of the model estimation. Finally, using the model estimates as a baseline, \autoref{sec:counterfactual} analyzes the outcomes of counterfactual experiments on different information and allocation policies across different measures of efficiency and welfare. 

\section{Institutional Background and Data}\label{sec:inst-data}

\subsection{Institutional Background}\label{sec:institution}
An organ procurement organization (OPO) is a non-profit agency responsible for recovering organs from deceased donors for transplantation. In the United States, there are 56 such agencies, each legally permitted to operate within an assigned donation service area. \OL\ from which we obtained data is one of the largest in the U.S., managing organ and tissue recovery for transplantation and serving  215 hospitals and 9 transplant centers in a region with a population of 20 million. In assigning organs, this %\OL\ 
OPO implements the assignment rules outlined by the Organ Procurement and Transplantation Network (henceforth, OPTN), which we will discuss next.

\paragraph{OPTN Priority Assignment} %
When an organ becomes available, the OPTN ranks all potential compatible recipients (patients) according to a priority assignment policy. Each patient’s priority is represented by their \emph{sequence number}, with those holding smaller numbers receiving offers before those with larger numbers. We refer to the process of sequentially offering an organ to this list of patients as a \emph{match run}. According to \cite{OPTN_2019}, the sequence number assigned to a patient in need of a lung transplant is determined by four factors: (i) the lung allocation score (LAS), which ranges from 0 to 100 and is calculated based on a combination of waitlist urgency and post-transplant survival probabilities; (ii) blood-type compatibility with the donor; (iii) total waiting time; and (iv) physical proximity to the donor. Initially, matched donor-patient pairs are grouped based on physical proximity and blood type compatibility. Within each group, priority is first given to patients with higher LAS, and if there is a tie, priority is then given to those who have waited longer.

\paragraph{The Allocation System} \label{sec:allocation}
As mentioned before, the OPTN system offers the organ to patients in ascending order of their sequence number in the match run. 
The start of the match run initiates a two-stage decision process involving the patient and their medical team, which we explain next.

The first stage is known as ``provisional yes,'' where patients (together with their medical team) need to decide within an hour, whether to \emph{provisionally} accept or reject the offer. Patients who decide to provisionally reject the offer are no longer eligible for that organ.\footnote{According to OPTN, the provisional yes practice informs the acting organ procurement agency that the patient (and their medical team) has received the offer and wishes to proceed, either by accepting the organ or seeking more information  \citep{OPTN_2019}. Introduced in 2018, this practice addresses concerns about patients who are not fit for transplantation but still appear on the waiting list, causing delays in the allocation process. By allowing these patients to opt out, it streamlines and expedites organ allocation by maintaining an active list of potential recipients.} After the first stage, the allocation process moves to the second stage, where patients who provisionally accepted the offer can then decide whether to \emph{finally} accept or reject it. At this stage, patients are informed of their sequence numbers and the reasons for any prior rejections, if applicable \citep{harhay2019donor}. If all final decisions result in rejection, then the organ will not be allocated.\footnote{Sometimes, a third stage of ``expedited placements'' may occur when the OPO faces a last-minute decline of a previously accepted organ or believes an organ to be viable but has been declined. In such cases, the OPO may make a direct offer to a center with a history of more liberal acceptance, regardless of the patient's position on the match run. This stage is more common in kidney allocations but occurs only occasionally with other organs. We abstract away from this process as we do not observe it in our data.} Importantly, neither provisional nor final rejections adversely affect patients' priority in future assignments.

When making the final acceptance decision, patients with lower priority (i.e., those with larger sequence numbers) should engage in the following contingent reasoning: they infer that, by the time it is their turn to decide, all higher-priority patients who provisionally accepted the organ must have subsequently rejected it. In other words, their decision to accept is contingent on previous rejections. This is because, if any of these higher-priority patients had accepted the offer, it would not be their turn to decide. Consequently, if previous rejections following provisional acceptance are informative about organ quality, this reasoning creates a channel for social learning. In Section \ref{sec:reduced_form}, we present evidence supporting this reasoning and highlights the role of social learning.

\subsection{Data}\label{sec:data}
Our data includes all lung donations managed by %\OL\ 
the concerned OPO between January 2019 and December 2020. For each donor organ, we observe a rich set of donor and patient characteristics, as well as 
detailed records of their acceptance and rejection decisions.\footnote{This includes the timing of decisions and the recorded reason for rejections.} The only key missing variable is the total waiting time of a patient at the time of receiving the offer. We construct this variable using data from the Scientific Registry of Transplant Recipients, which includes the initial wait-listing date for each patient.\footnote{The dataset from the Scientific Registry of Transplant Recipients (SRTR)  includes data on all donors, wait-listed candidates, and transplant recipients in the US, submitted by members of the OPTN. The Health Resources and Services Administration (HRSA) of the U.S. Department of Health and Human Services provides oversight to the activities of both the OPTN and SRTR contractors.} This allows us to calculate the patient's total waiting time for each offer by taking the difference between the date of the offer and their initial wait-listing date, excluding any inactive periods.%

To keep our empirical model tractable, we focus on patients over 12 years old who need a bilateral lung transplant.\footnote{This excludes patients who require both heart and lung transplants or a single lung transplant, as the challenges they face and the urgency of their conditions differ substantially from those needing a bilateral lung transplant. Patients under 12 are also excluded as they are prioritized differently by the OPTN.} 
Our dataset includes 32,786 offers, involving 548 deceased donors and 1,348 patients actively wait-listed during our observational period. Although all patients in our  dataset require a bilateral lung transplant, a small portion of them receive a single lung transplant; we observe about 5\% of accepted donors (out of 290 in total) had their lungs allocated to two different patients. In these cases, we treat the two allocations as a single allocation and use the smaller sequence number of the two patients as the accepted sequence number.

The high rate of unallocated organs is a striking feature of our data. With approximately 1,300 patients in need of a bilateral lung transplant and only 548 donors available, demand is more than double the supply. Despite this, the final allocation rate for these organs is only about 53\%. Specifically, out of the 548 deceased donors, about 258 had their lungs rejected by all matched patients. %
This substantial gap between available organs and their utilization underscores the importance of evaluating the efficiency and welfare outcomes of the current priority policy and exploring alternative policies, which we will discuss in \autoref{sec:counterfactual}.

\paragraph{Patient and Donor Summary Statistics} \autoref{tab:summary-stats} summarizes patients' and donors' characteristics in our sample. Panel (A) of \autoref{tab:summary-stats} reports sample statistics for patient characteristics, including health indicators such as body mass index (BMI) and diabetic status, together with medical history, such as whether the patient had any prior lung transplant. On average, patients have waited just over three months before receiving their first lung transplant offer. The average LAS is approximately 43 points, with over 80\% of patients having LAS between 30 and 50. Recall that a higher LAS indicates a greater likelihood of benefiting from a lung transplant, and the OPTN allocation system prioritizes patients with higher LAS. 

During the sample period, patients received an average of 24 offers in total. However, the median number of offers is only about 6, suggesting that many patients received substantially fewer offers.  The positive skewness in the number of offers received is likely due to a small portion of patients with lower LAS remaining on the waitlist for extended periods, thereby receiving a large number of offers.

Panel (B) of \autoref{tab:summary-stats} reports the sample statistics for donor characteristics. On average, donors are 18 years younger than the patients. Approximately 20\% of donors have either a history of heavy alcohol consumption or a status for increased risk, suggesting considerable variability in the quality of donor lungs. Despite this variability, the overall health of donor lungs is good, as indicated by the variable ``P/F ratio,'' which averages around 420 and falls within the normal and healthy range of 400 to 500 measured at sea level.\footnote{The P/F ratio is defined as the ratio between the partial pressure of arterial oxygen (PaO2, measured in mmHg) and the fraction of inspired oxygen (FiO2, expressed as a decimal). A P/F ratio below 300 suggests some level of impaired lung function.} However, although each donor is offered to an average of 60 patients, fewer than 50\% of these donor lungs are ultimately accepted and allocated. 

\begin{table}[t!]
    \centering
    \caption{Patient and Donor Characteristics}
    \label{tab:summary-stats}
    \begin{tabular}{lcccc}
        \hline \hline
         & Mean & S.D. & Min & Max \\
        \hline
       \multicolumn{5}{c}{\emph{Panel A: Patient Characteristics} $(N=1,348)$}   \\ \hline 
       Age & 55.9  &  13.5    & 12 & 79 \\
        Female  & 51.7\% &  50.0\%     &     0 & 100 \\
        Caucasian   & 67.5\% & 46.9\% & 0 & 100\\ 
        Body Mass Index (BMI) & 25.5  &  4.5   &15  &37\\
        Diabetic patient & 16.8\% & 37.4\%   &0   &100\\
        Previous lung transplant & 3.6\% & 18.7\% & 0 & 1 \\
        Lung Allocation Score (LAS) & 43.3&14.3 & 5.9 & 95.4 \\
    % &3.4 & 7.5& 0& 87.1\\ 
        Waiting time (months) &3.4 & 7.5& 0& 87.1\\ 
        Number of offers per patient & 24.3& 42.7 & 1 & 347 \\

       % \hline
      %  Number of observations & \multicolumn{4}{c}{$N = 1,348$} \\
       \hline
       
       \multicolumn{5}{c}{ \emph{Panel B: Donor characteristics} $(N=548)$} \\ \hline  
        Age  & 38.2 & 14.1  & 7 &75\\
       Weight (kg) & 79.0& 20.8 & 23.5 &      170.7 \\
       Height (cm) &169.5 &  10.8  &      119     & 198.5 \\
       Heavy alcohol history &17.7\% &38.2\%& 0 &1  \\
      IV drug use history &6.0\%&23.8\%& 0&1  \\
       Increasing risk status & 21.2\%& 40.9\%&0&1 \\
      P/F ratio  &   419.3 &   102.3   &   80  &   1400 \\   
      Number of offers per donor & 59.8 & 48.4 & 1 & 222 \\
  \hline \hline \\  [-3ex] 
    \end{tabular}
  \begin{minipage}{11cm} ~\\
    \small {\emph Note:} For all patient characteristics that can vary throughout our observational period, such as age, BMI, LAS, and waiting time, we use the value recorded at the first offer that each patient receives. 
  \end{minipage}
\end{table}

To see if there are any differences in lung quality between accepted and rejected donors, \autoref{tab:donor} reports summary statistics of donor characteristics by their final allocation outcome. As expected, accepted donors generally exhibit higher quality compared to rejected donors. For example, accepted donors are, on average, three years younger and have slightly lower probabilities of a history of heavy alcohol consumption or intravenous drug use compared to rejected donors. More importantly, the direct measure of lung quality, the P/F ratio, is approximately 10\% higher in accepted donors than in rejected ones, indicating that the former has better lung functionality than the latter. We also note that lungs from accepted donors are offered to approximately 20 more patients on average than rejected donors.

\begin{table}[t!]
    \centering
    \caption{Donor Characteristics by Allocation Outcome}
    \begin{tabular}{lcccc}
       \hline \hline
      &    \multicolumn{2}{c}{Accepted Donors} & \multicolumn{2}{c}{Rejected Donors} \\
        & Mean & S.D. & Mean & S.D.  \\
       \hline
        Age (year)  & 36.9 
        &13.4 & 39.7& 14.7\\
%Female &34.5\% & 47.6\%& 35.2\%& 47.8\%& 33.7\%& 47.4\%\\
        Weight (kg)   & 80.1 &      20.3 & 77.8 & 21.3\\
        Height (cm)    & 169.5     & 10.6 & 169.4& 11.1\\
        History of heavy alcohol consumption&  17.2\% &37.8\%  &18.2\%& 38.7\%\\
        History of IV drug use  & 5.5\%& 22.9\%  & 6.6\%& 24.9\% \\
        Increasing risk status  &21.4\%& 41.1\% & 20.9\%& 40.8\% \\
        P/F ratio     &  437.2  &   83.0 & 399.2& 117.4 \\
        Number of offers per donor  & 68.5& 48.2 & 50.1& 46.9\\
     %  \textbf{Accepting patient: position in queue} & 10.1 & 21.2  & -  & -\\
        \hline 
        Number of observations  & \multicolumn{2}{c}{$N=290$} & \multicolumn{2}{c}{$N=258$}\\
        \hline \hline \\ [-3ex]
    \end{tabular}
    \label{tab:donor}
     \begin{minipage}{12cm} ~\\
% \small {\emph Note: \textbf{increasing risk status?}}  
  \end{minipage}
\end{table}

\paragraph{Acceptance Rate and Offers} As described in \autoref{sec:institution}, patients first decide whether to provisionally accept the offer before making a final decision. So the observed low final acceptance rate\textemdash only about 53\%\textemdash could stem from either a high provisional rejection rate or a situation in which many donors are provisionally accepted but ultimately rejected. Our data suggests that the latter is the case: more than 95\% of all offers are provisionally accepted, yet less than 1\% result in final acceptance. This low final acceptance rate is underestimated as our dataset only includes final decisions up to the first acceptance in each match run. In other words, we do not observe the potential final decisions of patients with lower priority than the position at which the organ was accepted. Therefore, when considering the final acceptance decisions, we exclude all offers made after the first acceptance, leaving us with 15,773 offers for which both provisional and final decisions are recorded.

\begin{table}[t!]
    \centering
    \caption{Offers Characteristics by Decision Stage}
    \label{tab:summary-stats2}
    \begin{tabular}{lcccccc}
        \hline \hline
         &  \multicolumn{2}{c}{All} &  \multicolumn{2}{c}{Provisional Yes} &  \multicolumn{2}{c}{Final Yes}   \\
         & Mean & S.D. & Mean & S.D. & Mean & S.D.\\ 
        \hline    
        Lung allocation score (LAS) & 39.74 & 10.13 & 39.68 & 10.03 & 54.47 & 20.27 \\  
        Primary blood type match (\%) & 76.40 & 42.47 & 76.43 & 42.44 &  96.90 & 17.37  \\
         Waiting time (month) & 8.46 & 11.25 & 8.45 & 11.20 & 5.84 & 9.98 \\
        Distance (per NM) & 247.95 & 212.58 & 239.80 & 186.84 &  119.00  & 130.79 \\ 
       Decision time (minute) & - & - &   10.28 & 91.68 & 207.48 & 442.71\\ 
        \hline
        Number of observations  &  \multicolumn{2}{c}{$N= 32,786$} &   \multicolumn{2}{c}{$N= 31,329$} &  \multicolumn{2}{c}{$N=290$}\\
        \hline \hline \\  [-3ex] 
    \end{tabular}
    \begin{minipage}{15cm}   ~\\
     \small \emph{Note:} The variable ``primary blood type match'' is a categorical variable with three levels: 0 for incompatible, 1 for moderate compatibility and 2 for the highest compatibility. All observations in the dataset have a primary blood type match rating of at least 1. In addition, the unit for waiting time is in months, and the unit for distance is measured in nautical miles.
    \end{minipage} 
\end{table}
Table~\ref{tab:summary-stats2} reports the key characteristics of offers at different decision stages, shedding light on the acceptance decisions from the patients' perspective. Since most offers are provisionally accepted, the characteristics of all offers and those provisionally accepted are very similar. On average, an offer is provisionally accepted by a patient with a LAS of about 40, who has been waiting for approximately 8 months, has about a 76\% chance of being a primary blood type match with the donor and is located within 240 nautical miles of the donor.\footnote{Donor-patient pairs that are of primary blood type match include all four identical blood type combinations (O-O, A-A, B-B, AB-AB), as well as specific cross-type matches: O donor with B patient, A donor with AB patient, and B donor with AB patient. Pairs that are of secondary blood type match consist of O donor with a patient with blood type A or AB. All offers in our sample are either primary or secondary blood type matches.} The only noticeable difference between all offers and those provisionally accepted is the slightly shorter distance between the donor and the patient.

In terms of decision time, provisional decisions occur much faster than final decisions. On average, the former takes only about 10 minutes, whereas the latter requires an additional 3 hours. We also observe that offers that are ultimately accepted tend to be associated with patients who have higher LAS, are more likely to have a primary blood type match with the donor, are geographically closer to the donor, and have shorter waiting times. These relationships, except for waiting time, are as expected, as they all contribute to a smaller sequence number and therefore a higher priority for the patient in any match run, where the probability of acceptance is usually higher.

Our descriptive results suggest that patients treat the two stages of decision-making differently. Provisional decisions are made quickly and potentially influenced by factors such as the immediate availability of clinicians or the patient's physical conditions for transplantation. In contrast, final decisions require more deliberation on the quality of the match and thus take more time. Together, these observations motivate us to separately model these two decisions and the information involved in each (see \autoref{sec:model}).

\section{Reduced-form Evidence of Social Learning} \label{sec:reduced_form}
In this section, we present reduced-form evidence of social learning in lung transplant acceptance decisions. We show that, even after controlling for all variables that determine a patient's sequence number,  the probability of organ acceptance decreases as the sequence number increases.  
In particular, we find that (i) the four key variables\textemdash patient LAS, waiting time, and donor-patient blood type match and distance\textemdash affect the sequence number in a manner consistent with the description of the OPTN priority assignment, and (ii) sequence number exhibits a significant negative effect on final acceptance decisions, even after conditioning on all controls. These findings suggest that all else equal, a patient is less likely to accept an organ if they receive the offer at a higher sequence number. This suggests receiving an offer at a larger sequence number signals bad news about the offer. We also find that sequence number has no effect on provisional acceptance decisions once we condition on these key variables and other controls, consistent with the idea that patients treat provisional and final acceptance decisions differently.

\paragraph{Sequence Number} Recall that in \autoref{sec:inst-data}, we describe how sequence number is computed by the OPTN. Here, we present an alternative statistical model for assigning sequence numbers, examining how a patient's sequence number is related to their characteristics, donor characteristics, and donor-patient compatibility. This model serves as a control in the main reduced-form evidence for social learning.  Specifically, we estimate the following model:
\begin{equation}\label{eq:reducedform}
    s_{i, j} = \textbf{x}'_j \boldsymbol{\beta} + \textbf{x}'_i \boldsymbol{\gamma} + \textbf{c}'_{i,j} \boldsymbol{\alpha} + \eps_i, 
\end{equation}
where $s_{i, j}$ is the sequence number of patient $j$ when matched with donor $i$; $\textbf{x}_j$ is a vector of patient $j$'s characteristics that include their LAS, total waiting time, BMI, as well as health condition indicators such as diabetic status and history of previous lung transplant; $\textbf{x}_i$ is a vector of donor $i$'s characteristics including age, weight, height, P/F ratio and dummies for whether the donor has a history of IV drug or heavy alcohol use. We also include $\textbf{c}_{i,j}$, a vector of donor-patient compatibility indicators, such as primary blood type match, age, weight, and height differences, and the distance between donor $i$ and patient $j$.\footnote{More specifically, the geographical distance between the donor and the patient is characterized by ``Zones,'' where Zone A is for distances between 0-250 nautical miles (NM); Zone B is between 250-500 NM; Zone C is between 500-1000 NM; Zone D is between 1000-1500 NM; and Zone E is between 1500-2500 NM.} 

\begin{table}[t!]
  \centering
  \caption{Reduced-Form Analysis: Sequence Number, Provisional Acceptance and Final Acceptance}
   \label{tab:reduced_form}
  \begin{tabular}{lccc}
\hline \hline
   & (1) & (2) & (3) \\
      & Sequence number  & Provisional yes & Final yes \\ \hline 
   
    Sequence number &    & -$0.001$ &-$0.054^{***}$  \\
                    &    & (0.001)  & (0.010) \\ 
    Patient LAS &  -$0.596^{***}$ & -$0.018^{***}$ & $0.041^{***}$   \\
                    &  (0.016) & (0.003) &  (0.004)\\
    Primary blood type match  & -$39.938^{***}$ &  $0.020$ & $0.759^{*}$ \\
                    & (0.382)  &  (0.077) & (0.390)\\
    Waiting time (month) & -$0.064^{***}$  &  $0.005^{*}$ & -$0.022^{**}$\\
                    & (0.013)   &  (0.003) & (0.010)\\
   \multicolumn{2}{l}{Distance (reference group--Zone A: 0-250 NM)}  & &  \\
        ~    Zone B: 250-500 NM & $40.807^{***}$ & $0.922^{***}$ & -$0.888^{***}$ \\ 
                  & (0.313) & (0.078)  & (0.156)\\
        ~    Zone C: 500-1000 NM & $31.356^{***}$ & -$1.739^{***}$  & -$1.810^{*}$\\ 
                  & (1.770) & (0.094)  & (1.025)\\
        ~    Zone D: 1000-1500 NM & $23.893^{***}$& -$1.180^{***}$  & - \\ 
                  & (2.892) & (0.234)  & - \\
        ~    Zone E: 1500-2500 NM & $18.203^{***}$& -$2.485^{***}$  & - \\ 
                  & (2.419) & (0.176)  & - \\
   % Female patient &  $0.512$  &  -$0.109$ \\
     %               & (0.672) & (0.153)  \\
  %  Donor age & 0.006 & $0.006$ & -$0.027$ \\
  %                  & (0.007) & (0.007) &  (0.031) \\
  %  Donor weight & -$0.014^{**}$ & -$0.014^{**}$ & -$0.000$ \\
%                & (0.006) & (0.006) & (0.028) \\
%    Donor height & $0.037^{***}$ & $0.036^{***}$ &  -$0.161^{**}$\\
%                & (0.011) & (0.011) &  (0.049) \\
%    Donor drug & $0.075$ &  -$0.029$ & -$8.254^{***}$ \\
 %                & (0.277) &  (0.296)&  (1.065) \\
%   Donor alcohol & -$0.044$ & -$0.046$ &  $9.119^{***}$\\
%                 & (0.171) &  (0.175) & (0.756) \\
%    Donor risk & -$0.107$ &   -$0.143$ & $ 4.502^{***}$\\
%                &(0.157) &  (0.161)  & (0.694) \\
%    Weight difference &  -$0.014^{***}$ &  -$0.014^{***}$ & -$0.050^{**}$ \\
 %               & (0.005) & (0.005) &  (0.024)\\
 %   Age difference & $0.022^{***}$ & $0.022^{***}$ & $0.114^{***}$\\
 %               & (0.005) & (0.005) &  (0.022)\\
 %   Height difference &  $0.039^{***}$ &  $0.037^{***}$ & $0.317^{***}$\\
  %              & (0.009) &   (0.010)&  (0.044)\\
    Patient characteristics & X & X & X \\
    Donor characteristics & X & X & X \\
    Patient-donor differences & X & X & X \\
    Provisional yes & - & - & X \\    
    \hline
    Observations &  32,697 & 32,697 &  14,587 \\
    R-squared &  0.490  & 0.125  & 0.246\\     R-squared w/o $s_{i,j}$ &  -  & 0.125  & 0.207\\            
  \hline \hline \\  [-3ex] 
  \end{tabular}
  \begin{minipage}{14cm} ~\\
    \small {\emph Note:} For all regressions, we control for a wide range of variables about the patient, donor and donor-patient characteristics which we do not report due to space constraints.  Note that since we restrict the sample to offers that are provisionally accepted in Column (3), the variable ``sequence number'' is reordered based on the initial sequence number of these patients. Robust standard errors are reported in parentheses.
    $^{***}$Significant at 1\% level, $^{**}$Significant at 5\% level, $^{*}$Significant at 10\% level. Columns (2)-(3) show results from a logit regression, so pseudo R-squared is reported. %
  \end{minipage}%
\end{table}

The first column of \autoref{tab:reduced_form} reports the results of regression~\eqref{eq:reducedform}. The coefficients of interest are those that correspond to the four factors used by the OPTN to assign sequence numbers: patient LAS, waiting time, and donor-patient blood type match and distance. All four coefficients are statistically significant and exhibit the anticipated signs: patients with higher LAS, a better blood type match, longer waiting times, and closer geographical proximity to the donor are more likely to have lower sequence numbers; thus receiving higher priority in the queue. In terms of magnitude, the coefficients for blood type match and distance are much larger than those for LAS and waiting time, suggesting their first-order effects in determining priority levels, which is consistent with the description in Section \ref{sec:allocation}.

\paragraph{Acceptance Decisions}
Next, we examine the effect of sequence number on patient acceptance decisions, controlling for all previous variables, including those used by the OPTN to determine the sequence number. To do so, we run a Logit model as follows:
\begin{equation} \label{eq:accept}
y_i = \mathbbm{1}\{\beta_s 
s_{i,j} + \textbf{x}'_j \boldsymbol{\beta} + \textbf{x}'_i \boldsymbol{\gamma} + \textbf{c}'_{i, j} \boldsymbol{\alpha} + \eps_i \geq 0\},
\end{equation}
where, as before, $s_{i, j}$ is the sequence number of patient $j$ when matched with donor $i$, and the vectors \textbf{x}$_j$, \textbf{x}$_i$ and $\textbf{c}_{i, j}$ represent the observed characteristics of patients, donors, and donor-patient compatibility. In the specifications below, $y_i$ is a dummy variable indicating whether the donor is provisionally accepted or finally accepted.

\paragraph{Provisional Yes} The second column of \autoref{tab:reduced_form} presents the result of regression~\eqref{eq:accept} when $y_i$ represents the provisional acceptance decision. As reported in the first row, after controlling for a rich set of observables, the negative coefficient $\beta_s$ on sequence number is not statistically significant. This suggests that sequence number itself does not affect patients' provisional decisions beyond the information included in the controls. The coefficient on patient LAS is negative and statistically significant, indicating that patients with a higher LAS are more likely to opt out the allocation process. We also observe a weak positive coefficient for waiting time, suggesting that patients who have waited longer are slightly more likely to accept provisionally. Finally, compared to patients in Zone A (within 0-250 NM of the donor), only patients in Zone B (within 250-500 NM of the donor) are more likely to provisionally accept the offer, while those in more distant zones are less likely to do so. This is consistent with the idea that the provisional decisions reflect basic unavailability for a potential transplant operation.

\paragraph{Final Decisions} The third column of \autoref{tab:reduced_form} presents the result of regression \eqref{eq:accept} when $y_i$ represents the final acceptance decision. As mentioned before, our data on patient final acceptance is truncated at the first acceptance.\footnote{ Recall that we exclude all offers following the first acceptance, resulting in 15,773 offers where both provisional and final decisions are recorded.} The coefficient of interest here is the one associated with the sequence number. Unlike provisional acceptance decisions, this coefficient is negative ($\hat{\beta}_s = -0.054$) and statistically significant at the 1 percent level, even after controlling for all factors used to determine sequence numbers, as well as all other observables. 

In other words, all else equal, having a larger sequence number alone decreases the probability of a patient ultimately receiving a transplant. This significant negative correlation between sequence number and final acceptance probability suggests that transplant centers use patients' sequence numbers to make negative inferences about donor quality. Since a donor can only be allocated to the next patient in line if it is rejected by all higher-priority patients, only information about rejections is transmitted as the offer moves down the priority queue. The negative coefficient on sequence number implies that previous rejections are interpreted as ``bad news'' about the donor quality, thereby reducing the likelihood that the donor will finally be accepted.

We note that all coefficients for patient LAS, waiting time, and distance are statistically significant in final acceptance decisions but have opposite signs compared to those in provisional acceptance decisions. For example, the coefficient for LAS changes from negative in column (2) to positive in column (3) of \autoref{tab:reduced_form}. This suggests that, unlike in provisional decisions, patients with a higher LAS\textemdash indicating a more urgent status and better survival probability\textemdash are more likely to receive a transplant, despite having a lower likelihood of provisional acceptance. Similarly, a negative coefficient for Zone B (compared to Zone A) implies that patients residing farther from the donor are less likely to receive a transplant, even though they have a higher likelihood of provisional acceptance. Motivated by these findings, we model these two decision stages separately in the following section.

\section{Structural Model}\label{sec:model}

We propose a model that combines a discrete-choice specification with sequential decision-making which creates channels for social learning. To do so, we build on the classic theoretical model of sequential social learning \citep{banerjee1992simple, BichHirshWelch:92}, adapting it for the purposes of structural estimation. Specifically, our model incorporates four key components that influence a patient's acceptance decisions: (i) unobserved lung quality that is common to all matched patients; (ii) observed donor and patient individual characteristics; (iii) observed donor-patient match quality characteristics; (iv) unobserved heterogeneity. The first component, an unobserved common-value component, is the focus of social learning, where patients update their beliefs negatively about this common quality based on past rejections within a match run. The second and third components capture 
observed heterogeneity among patients, donors, and donor-patient pairs. 

\paragraph{Donor Quality} Each donor $i\in N$ is endowed with a quality $\theta_i\in\{-1,1\}$, which is identically and independently distributed across donors. We denote by $p\in(0,1)$ the prior probability that the donor quality is $\theta=1$. Throughout, we refer to $\theta_i = 1$ as donor $i$ being of high quality and $\theta_i = -1$ as donor $i$ being of low quality. Each $\theta_i$ captures any unobserved common aspects of the donor's lung quality, in addition to the observed donor characteristics, and it 
is unobserved by both the patients and the econometrician. We assume that it is fixed
throughout the allocation process. This assumption is supported by our empirical
observation that nearly all offers\textemdash about 98\% in our dataset\textemdash
receive a final decision before the organ is extracted, if at all, from the donor.

\paragraph{Allocation} When donor $i$ becomes available, a match run is initiated for a set of compatible patients, denoted by $N_i$. These patients, indexed by $j$, first make a provisional decision; those who respond with a provisional yes are then asked for a final decision. Provisional Yes decisions are made based on idiosyncratic factors such as the availability of a transplant team at the center and the patient's readiness for transplantation. In contrast, final acceptance decisions are based on the patient's assessment of donor-patient compatibility and the quality of the donor organ. 

Formally, we assume that each patient $j \in N_i$ is endowed with two signals, denoted by $(\nu_{j},\omega_{i,j})$, each taking values in $\{-1,1\}$. The first signal $\nu_j$ indicates whether patient $j$ is ready for transplant ($\nu_{j}=1$) or not ($\nu_{j}=-1$), and is independent of the donor quality $\theta_i$. We assume that $\nu_{j}$ are i.i.d.\ across patients $j$, with $\Prob[\nu_j =1] = \mu\in (0,1)$. The second signal $\omega_{i,j}$ is informative about the donor quality, with precision $\alpha \in (1/2, 1)$, i.e., $\alpha = \Prob[\omega_{i,j} = k|\theta_i=k]$ for all $k \in \{ -1, +1\}$. We assume that conditional on $\theta_i$, $\omega_{i, j}$ are i.i.d.\ across patients $j$.

\paragraph{Final Acceptance} Denote by $P_i := \{j \in N_i: \nu_{j}=1\}$ the set of patients who are ready to receive a transplant from donor $i$. When offered the organ from donor $i$, patient $j\in P_i$  decides whether to accept ($d_{i,j}=A$) or reject ($d_{i,j}=R$) the organ.\footnote{By default, the decisions of patients in $N_i\setminus P_i$ are set to $R$.} Patient $j$'s utility from rejecting the organ is given by $\epsilon_{i,j}^R$. If patient $j$ accepts a donor of quality $\theta_i$, their utility is given by
\begin{equation} \label{eq:utility_patient}
    u^A(\tilde{\mathbf{x}}_{i, j}, \theta_i, \epsilon^A_{i,j}) = \Tilde{\mathbf{x}}'_{i,j} \boldsymbol{\tilde\beta}  +  \gamma \theta_i + \epsilon_{i, j}^A,
\end{equation}
where $\tilde{\mathbf{x}}_{i,j}':= (1, \mathbf{x}_i', \mathbf{x}'_{j}, \textbf{c}_{i,j}')$ consists of three observable components, including a constant that normalized to one: $\mathbf{x}_i$ is a vector of donor $i$'s characteristics; $\mathbf{x}_j$ is a vector of patient $j$'s characteristics; and $\textbf{c}_{i, j}$ is a vector of characteristics capturing the compatibility between donor $i$ and patient $j$ (just as in the reduced-form model in \autoref{sec:reduced_form}). The variables in $\tilde{\mathbf{x}}_{i, j}$ are observable to both patient $j$ and the econometrician. The parameter $\boldsymbol{\tilde\beta}$ is a vector consisting of the utility weights associated with $\tilde{\mathbf{x}}_{i, j}$ and $\gamma$ is the utility weight associated with $\theta_i$. Finally, $\epsilon^A_{i,j}$ denotes the idiosyncratic utility shock that patient $j$ receives when accepting donor $i$'s lung. We assume that $\epsilon^R_{i, j}$ and $\epsilon^A_{i, j}$ are distributed i.i.d.\ across patients and donors, each following a Type I extreme value distribution. Consequently, the difference between them, denoted by $\delta_{i, j}:= \epsilon^A_{i,j} - \epsilon^R_{i,j}$, follows a logistic distribution centered at zero with a scale parameter of one.

Before deciding whether to accept or reject donor $i$, patient $j \in P_i$ observes the idiosyncratic $(\eps^R_{i, j}, \eps^A_{i, j})$, the informative signal $\omega_{i, j}$, and their sequence number $s_{i,j}$.\footnote{Note that both the patients' idiosyncratic shocks and the informative signals are unobserved by other patients and the econometrician.} Based on this information, patient $j$ accepts the offer if the expected  utility from accepting is greater than that from rejecting:
\[ 
\mathbb{E}[u^A(\Tilde{\mathbf{x}}_{i,j}, \theta_i, \epsilon^A_{i, j}) -u^R(\epsilon^R_{i, j})| \omega_{i, j}, s_{i, j}] \geq 0.
\]
From the perspective of patient $j$, the only unknown variable in the above expression is $\theta_i$. Thus, we can write the above inequality as 
\begin{equation}\label{eq:accept_threshold}
\delta_{i, j} \geq - v(\Tilde{\mathbf{x}}_{i, j}, \omega_{i, j}, s_{i, j}),
\end{equation}
where 
\begin{equation} \label{eq:accept_utility}
    v(\Tilde{\mathbf{x}}_{i, j}, \omega_{i, j}, s_{i, j}) = \Tilde{\mathbf{x}}'_{i,j} \boldsymbol{\tilde\beta} + \gamma \mathbb{E}[\theta_i|\omega_{i, j}, s_{i, j}].
\end{equation} 
Thus, \eqref{eq:accept_utility} represents patient $j$'s final acceptance threshold for donor $i$, which is determined by a combination of donor and patient characteristics (including those describing their mutual compatibility), the patient's private information, and the information conveyed through the allocation process via the patient's sequence number. The knowledge of patients' sequence number plays an important role: when evaluating the donor’s 
quality, a patient should condition their evaluation on the event that all other patients with smaller sequence numbers must have rejected the offer; otherwise, the offer would not have reached them in the first place. This leads to a problem of negative inference as the donor is offered further down the match run, which we discuss next.

\paragraph{The Inference Problem Conditional on Provisional Yes} 
Since all patients must provisionally accept the offer before making a final decision, patient $j$ infers, when it is their turn, that all higher-priority patients have provisionally accepted but ultimately rejected the offer. According to \eqref{eq:accept_threshold}, these final rejections are influenced by a combination of private information and patient characteristics. To make the inference problem tractable, we assume that each patient $j \in P_i$ knows the characteristics of other patients in $P_i$ with smaller sequence numbers.\footnote{This assumption also help ease the computational complexity of estimating our structural model.}  

Given these patient characteristics, patient $j$ can thus infer the extent to which past rejections are 
informative about donor quality. Recall that the expected quality of $\theta_i$ conditioned on $\omega_{i, j}$ and $s_{i,j}$ is $$
     \mathbb{E}[\theta_i|\omega_{i,j}, s_{i,j}] = \mathbb{E}[\theta_i|\omega_{i,j}, (d_{i,j'})_{j' \in P_i, s_{i, j'} < s_{i, j}}=R] .$$
As $s_{i, j}$ increases, the number of patients in $P_i$ who ultimately reject donor $i$ also weakly increases. Consequently, one would expect  $\mathbb{E}[\theta_i|\omega_{i, j}, s_{i,j}]$ to be weakly decreasing in $s_{i, j}$. Intuitively, as the sequence number increases, the number of prior rejections increases, leading to an accumulation of negative information about donor quality. This reinforces doubt about the donor, further reducing the probability of acceptance and triggering a cascading effect of rejections. 

Thus, the amount of negative information conveyed through the sequence number can be seen as a measure of the extent of social learning in the current OPTN system. In \autoref{sec:counterfactual}, we conduct counterfactual experiments to assess the impact of social learning by varying the amount of information in the sequence number, ranging from no information to the maximum amount of private information. Taking into account the effect of social learning, we then examine how different priority assignment algorithms influence the informational content of the sequence number and their welfare implications.

\section{Estimation}\label{sec:estimation}%
Motivated by our empirical evidence (see \autoref{sec:reduced_form}), we estimate our model using maximum likelihood in two steps. First, from the data we directly estimate the probability of provisional acceptance $\mu$. Second, we estimate the remaining parameters using the Nelder-Mead simplex algorithm, a direct search method. The parameters estimated in the second step include: $\alpha$, the precision of the informative private signals; $p$, the prior probability that the lung quality is high; $\gamma$, the coefficient for the expected lung quality; and $\boldsymbol{\tilde\beta}$, the vector of coefficients for donor, patient and donor-patient match quality characteristics. Conditional on all provisional acceptances\textemdash which occurs with probability $\mu$\textemdash the vector of parameters $\Phi:= (\alpha, p, \gamma, \boldsymbol{\tilde\beta})$ collectively governs the final decision process. 

For each donor $i$ and each patient $j \in N_i$, we observe the following set of variables in the data: 
\[
\Lambda_{i, j} = (\mathbbm{1}_{j \in P_i}, \mathbbm{1}_{d_{i,j}= A}, s_{i, j}, \textbf{x}_i, \textbf{x}_j, \textbf{c}_{i,j})
\] 
where $\mathbbm{1}_{j \in P_i}$ is an indicator for whether patient $j$ provisionally accepts donor $i$; $ \mathbbm{1}_{d_{i,j}= A}$ is an indicator for whether patient $j$ finally accepts donor $i$; $s_{i,j}$ is the sequence number assigned to patient $j$ for donor $i$; $\textbf{x}_i$ is a vector of donor $i$'s characteristics; $\textbf{x}_j$ is a vector of patient $j$'s characteristics; and $\textbf{c}_{i,j}$ is a vector of donor-patient compatibility characteristics. Let $\Lambda_i = (\Lambda_{i, j})_{j \in N_i}$ denote the data observed for donor $i$, and $\Lambda = (\Lambda_i)_{i \in N}$ denote the data observed for all donors.

\subsection{The Likelihood Function} 
In the first step, since provisional decisions are governed by the patient's first signal $\nu_j$, which are i.i.d.\ across patients and independent of donor quality, the maximum likelihood estimator of $\mu$ is given by $\frac{\sum_{i \in N}|P_i|}{\sum_{i \in N}|N_i|}$. In the second step, given a vector of parameters $\Phi$, we first focus on the probability that donor $i$ is accepted by some patient $j \in P_i$. According to \eqref{eq:accept_threshold}, the probability of patient $j$ accepting donor $i$ is 
\[
\Prob[d_{i,j} = A|\tilde{\textbf{x}}_{i, j}, \omega_{i, j}, s_{i, j}; \Phi] = 1-F(- v(\Tilde{\textbf{x}}_{i, j}, \omega_{i, j}, s_{i, j}; \Phi)),
\]
where $F$ is the cumulative distribution function of the standard logistic distribution. The complementary probability, $F(- v(\Tilde{\textbf{x}}_{i, j}, \omega_{i, j}, s_{i, j}; \Phi))$ is thus the probability of patient $j \in P_i$ rejecting donor $i$. Denote by $$j^{*}_i= \argmin_{s_{i, j}}\{j \in P_i: d_{i, j} = A\}$$ 
the patient in $P_i$ who finally accepts donor $i$ and has the smallest sequence number. So, if there are multiple $j \in P_i$ such that $d_{i, j}= A$, donor $i$ is allocated to patient $j^*_i$.  

Since patients' idiosyncratic shocks are i.i.d., the likelihood that donor $i$ is finally accepted by $j^*_i$ under  $\Phi$ is 
\begin{align*}
\mathcal{L}^A_{i}(\Phi; \Lambda_i) = \big(1-F(- v(\Tilde{\textbf{x}}_{i, j^*_i}, \omega_{i, j^*_i}, s^i_{j^*_i}; \Phi)) \big)\cdot \prod_{j \in P_i: s_{i, j} < s^i_{j^*_i}} 
 F(- v(\Tilde{\textbf{x}}_{i, j}, \omega^2_{i, j}, s_{i, j}; \Phi)).
\end{align*}
Note that the first term is the probability that donor $i$ is accepted by patient $j_i^*$, \emph{conditional} on  all higher-priority patients rejecting the offer despite provisionally accepting it. The second term is the probability of this conditional event\textemdash i.e., that donor $i$ is ultimately rejected by these higher-priority patients who had provisionally accepted the offer. In contrast, the probability of donor $i$ being rejected by all patients in $P_i$ is \[
\mathcal{L}^R_i(\Phi; \Lambda_i) = \prod_{j \in P_i} F(- v(\Tilde{\textbf{x}}_{i, j}, \omega_{i, j}, s_{i, j}; \Phi)).\]
Let $S \subseteq N$ denote the set of donors that are successfully allocated, i.e., it has been accepted by at least one patient in $P_i$. The log-likelihood function of the observed donor allocations is thus equal to 
\begin{align}\label{eq:LL}
   \mathcal{LL}(\Phi; \Lambda):= \sum_{i \in S} \log \mathcal{L}^A_{i}(\Phi; \Lambda_i ) + \sum_{i \in N \setminus S}  \log \mathcal{L}^R_i(\Phi; \Lambda_i).
\end{align}
The first term in \eqref{eq:LL} is the log-likelihood of all donors in $S$ being accepted, while the second term is the log-likelihood of all donors in $N \setminus S$ not being accepted by any matched patients. %

\subsection{Estimation Results}
\begin{table}[h!] 
    \centering \caption{Estimation Results}
    \begin{tabular}{lcc}
    \hline \hline
      Parameter   & Estimate  & S.D. \\
      \hline
     $\mu$      & 0.958 & --\\
      $\alpha$   & 0.850 & 0.039 \\
      $p$   & 0.383 & 0.046 \\
      $\gamma$  & 4.934 & 0.005 \\
      $\boldsymbol{\tilde{\beta}}$ (selected) & & \\
\quad   Patient LAS & 0.033 & 0.001 \\
  \quad Patient waiting time (month) &  -0.024& 0.002 \\
  \quad  Donor-patient primary blood type match & 0.775 & 0.006
 \\  
  \quad  Donor-patient distance (per NM) & -0.003 & 0.001\\
         \hline \hline \\ [-3ex] 
          \end{tabular}
    \label{tab:estimation} 
  \begin{minipage}{11cm} ~\\
    \small {\emph Note: The standard deviations are obtained through the BHHH estimator.}  
  \end{minipage}
\end{table}
We report the coefficient estimates and corresponding standard deviations for selected parameters in \eqref{eq:LL} in Table~\ref{tab:estimation}. The full estimation results are provided in Table \ref{tab:estimation_full} in the appendix. Recall that $\mu$ is the provisional acceptance probability, which reflects the overall readiness of the patient pool to receive a transplant. Our estimate of 0.96 suggests that provisional acceptances are extremely common among lung transplant patients. Additionally, $\alpha$ represents the precision of the patients' private information regarding the unobserved quality of the donor's lungs, and $p$ represents the patients' prior belief that the quality is high. Our estimates for $\alpha$ and $p$\textemdash 0.85 and 0.38, respectively\textemdash suggest that although patients generally hold a pessimistic view of the unobserved quality, the private information they receive is highly informative.

The estimated coefficients $\boldsymbol{\tilde\beta}$ for the patient, donor, and donor-patient compatibility characteristics  show the expected signs, have reasonable magnitudes, and align with our reduced-form estimates. For example, a patient's utility for accepting a donor increases with a higher LAS or a primary blood type match, while it decreases as the patient waits longer or as the geographical distance between the patient and donor increases. Notably, the estimated coefficient for unobserved donor quality, $\gamma$, is about five times larger than that for a primary blood type match, which is the largest coefficient among all patient, donor, and donor-patient characteristics. This highlights the critical role of unobserved donor quality in determining patient acceptance decisions.

 \begin{figure}[t]
    \centering    
     \caption{Conditional Acceptance Probability by Sequence Number}
     \includegraphics[scale=0.37]{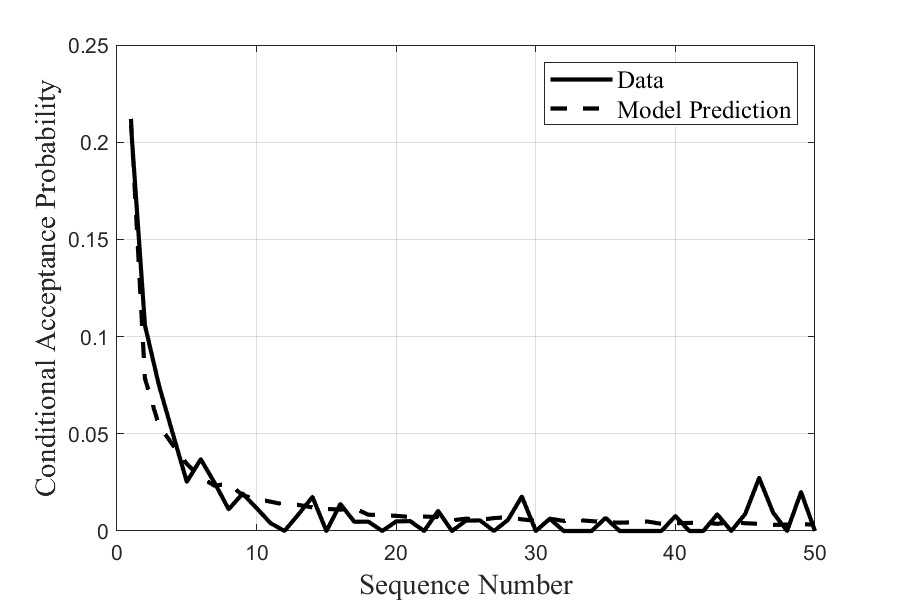}
    \label{fig:modelfit}
\end{figure}
To evaluate our model fit, we use these estimates to obtain predicted acceptance probabilities for patients at different sequence numbers. Figure~\ref{fig:modelfit} plots the average predicted acceptance probabilities and compares them with the empirical acceptance rate by sequence number. As shown in the figure, our model generally fits well with the observed pattern of acceptance, particularly in capturing the sharp decline in the acceptance rate at the initial sequence numbers. For instance, the acceptance rate at sequence number one in the data is approximately 21 percent, dropping by half to around 11 percent at sequence number two\textemdash closely aligning with our model predictions of 21 and 8 percent, respectively. 

Note that all reported probabilities, both predicted and empirical, are conditional acceptance probabilities. For example, if the acceptance probability at sequence number 5 is 0.1, this indicates a 10\% chance that the offer will be accepted by the patient who is fifth in line, conditional on all patients with sequence numbers 1 to 4 having rejected the offer.  Using these conditional probabilities, we can calculate the likelihood that the organ is allocated by sequence number, say, 50.\footnote{Formally, let $x=(x_1, x_2, \ldots)$ be the vector of conditional acceptance probabilities where $x_i$ denotes the probability of accepting the offer at sequence number $i$. The probability that the offer is accepted by sequence number $n$ is $\sum_{k=1}^n (\prod_{i=1}^{k-1} (1-x_i) x_k)$. 
We limit the calculation to sequence number 50, as the conditional acceptance probabilities become negligible beyond this point.} The likelihoods derived from the predicted and empirical probabilities are about 57\% and 56\%, respectively\textemdash both close to the observed overall allocation rate of 53\% in the data.

\section{Counterfactual Experiment} \label{sec:counterfactual}
Using our structural estimates, we conduct a series of counterfactual experiments to examine the impact of social learning on allocation outcomes and its interaction with different allocation policies. As a baseline, we simulate the organ allocation process under the OPTN priority assignment policy described in \autoref{sec:inst-data}. We then consider two sets of counterfactual experiments: one modifies the priority assignments to evaluate the effects of different policy proposals, while the other adjusts the information available to patients to better understand the role of social learning.

To implement the counterfactual simulations, we first construct the simulated data. For each donor, we randomly draw a set of blood type-compatible patients from the entire patient pool, ensuring that the number of assigned 
patients matches the data. We then rank these patients according to different priority policies. In our baseline experiment, we replicate the OPTN's priority assignment algorithm. For alternative policies, we compute the predicted acceptance probabilities using our structural estimates and rank patients accordingly. This results in different patient samples for each priority policy. For each sample, we randomly draw donor quality, patients' signals, and their idiosyncratic shocks according to our model estimates. We then compute the expected donor quality based on the information available for each patient, which varies across informational treatments. This process leads to a final decision (acceptance or rejection) for each patient. If the donor is allocated, we calculate the utility for the patient receiving the offer according to \eqref{eq:utility_patient}.\footnote{Specifically, we calculate what is known as the ex-ante maximum expected utility. As is well known, the ex-ante maximum expected utility of patient $j$ before the realization of her idiosyncratic utility shocks $(\eps^A_{i, j}, \eps^R_{i, j})$ is 
\begin{equation*} 
\E[\max\{ u^A(\Tilde{\textbf{x}}_{i, j}, \theta_i, \eps^A_{i,j}), u^R(\eps^R_{i,j})\}] = c + \log\Big(1+\exp{\big(\Tilde{\textbf{x}}'_{i,j} \boldsymbol{\tilde\beta}  +  \gamma \theta_i\big)}\Big),  
\end{equation*}
where $c$ is Euler's constant, $c \approx 0.577$.}   

\subsection{Alternative Priority Policies}
This section compares two alternative priority policies to the one currently used by the OPTN. These two policies differ in how they rank patients in a match run. The first policy, called  \emph{greedy priority}, offers the organ to patients in descending order of their estimated acceptance utility. The second policy, called \emph{reverse greedy priority}, offers the organ to patients in ascending order of their estimated acceptance utility.

In the absence of social learning, greedy priority maximizes the speed of organ allocation and ensures that the organ goes to the patient who values it the most, as those with the largest acceptance utility also have the highest acceptance probability. Thus, if the sole concern is allocation efficiency or ensuring the organ is given to the highest-value patient, without considering social learning, greedy priority seems to be the best policy. This policy, however, is also the ``worst'' when the goal is to minimize the negative inference from social learning: when an organ is rejected by patients who are most likely to accept it, subsequent patients infer the most negative information, making them more likely to reject as well. Consequently, in the presence of social learning, the net effect of greedy priority on allocation efficiency and patient welfare remains unclear.

In contrast, reverse greedy priority minimizes the impact of social learning, as the negative inference from past rejections is least informative under this policy. This is because these rejections come from patients who were more likely to reject the organ in the first place. So this policy is the ``best'' when the goal is to minimize the negative effects of social learning. However, since it prioritizes patients with the lowest acceptance utility, it leads to the lowest acceptance utility and the slowest allocation process. In the rest of this section, we present empirical results examining the effects of both policies on allocation efficiency, patient welfare, and the extent of social learning, and compare them to the current OPTN priority policy.

\begin{table}[t]
    \centering
    \caption{Counterfactual 1: Different Priority Policies with Social Learning}
    \begin{tabular}{lccc}
    \hline\hline
   &  (1) & (2) & (3)   \\ 
  Priority Policy  & OPTN (Current) &  Greedy  & Reverse Greedy \\
    \hline
     \multicolumn{4}{c}{\emph{Panel (A): Efficiency Outcomes }}  \\
    Allocation Rate (\%)	& 47.22  & 46.26  &  49.33
    \\
    Accepted Sequence Number & 6.91	 & 5.13 & 15.49
   \\
   \hline
   \multicolumn{4}{c}{\emph{Panel (B): Welfare Outcomes }} \\
 %  Acceptance Utility (excl. $\theta$)
 %   & 0.83 & 0.90 &  0.63\\ 
 % Total Utility (excl. $\theta$)  &204.28 & 216.11 &  161.79
 % \\ 
    Acceptance Utility  &	2.52	&2.75 & 1.48\\
    Total Acceptance Utility %(incl. $\theta$) 	
    & 619.72	& 662.27  &  381.03\\ 
   Acceptance Rate of High Quality (\%)	& 32.63 &32.44 & 32.63
 \\
   Rejection Rate of Low Quality  (\%)  	& 44.53	&45.30 & 42.42
 \\
   \hline
 \multicolumn{4}{c}{\emph{Panel (C): Accepted Patient Characteristics}}   \\
 LAS  & 59.12	& 64.24 & 38.15
  \\
 Waiting time (month)  &5.35 &	3.03 & 8.27
  \\
Primary blood type match 	& 1.61 &	1.60 & 1.54
  \\
Distance (NM)  &	99.76	& 155.73 & 339.75
  \\ \hline
Provisional Yes & $\checkmark$  & $\checkmark$  & $\checkmark$  \\
Available Information  & \multicolumn{3}{c}{one private signal and past rejections}   \\
    \hline\hline
    \end{tabular}
    \label{tab:counter_efficiency}
    \begin{minipage}{15.5cm} ~\\
    \small {\emph Note: The variable ``Acceptance Utility'' represents the average utility of patients who accept the offer. %which is equal to  $\E[\max\{u(\Tilde{\textbf{x}}_{i, j}, \omega^2_{i, j}, s_{i, j})+ \eps^A_{i,j}, \eps^R_{i,j}\}]$ where  $u(\Tilde{\textbf{x}}_{i, j}, \omega^2_{i, j}, s_{i, j}) = \Tilde{\textbf{x}}'_{i,j} \boldsymbol{\tilde\beta} + \gamma\theta_i$. %The variable ``Acceptance Utility (incl. $\theta$)'' is the same object but with $v(\Tilde{\textbf{x}}_{i, j}, \omega^2_{i, j}, s_{i, j}) = \Tilde{\textbf{x}}'_{i,j} \boldsymbol{\tilde\beta} + \gamma \theta_i$. 
    The variable ``Total Acceptance Utility'' is the sum of all acceptance utilities. In cases where multiple patients accept the same donor, the donor is allocated to the patient with the highest priority according to the corresponding priority policy.}
  \end{minipage}
\end{table}

Panel (A) of \autoref{tab:counter_efficiency} presents two efficiency outcomes: the average allocation rate and the average accepted sequence number. Column (1) reports our baseline simulation which replicates the OPTN priority policy. We find that our simulation is consistent with what is observed in the data (see Section \ref{sec:data}). While the actual allocation rate is 53\%, with an average accepted sequence number of 8.5, our simulation yields an average allocation rate of 47\% and an average accepted sequence number of 7. Columns (2) and (3) present the outcomes from greedy and reverse greedy priorities. Reverse greedy increases the allocation rate by 2 percentage points compared to the OPTN policy, while greedy priority reduces it by about 1 percentage point. Thus, the shift between the ``best'' and the ``worst'' policy in terms of minimizing the effect of social learning results in an approximately 3 percentage point difference in the overall allocation rate. This increased allocation rate under reverse greedy does not come without a cost, as it results in the slowest allocation process, with an average accepted sequence number of around 15\textemdash about three times that under greedy priority. These results highlight a trade-off between maximizing the allocation rate and expediting the allocation process when evaluating different priority policies under social learning.

Panel (B) of \autoref{tab:counter_efficiency} reports welfare outcomes under different priority policies. In terms of patient welfare, we compute the average utility of patients who finally receive an organ. As expected, greedy priority achieves the highest average acceptance utility, followed by the current OPTN policy, and then reverse greedy priority. In terms of social welfare, greedy priority also achieves the highest total acceptance utility\textemdash the sum of the utilities of all patients who receive an organ\textemdash despite having the lowest allocation rate. Therefore, from the perspective of a social planner who aims to maximize total welfare and expedite the allocation process, greedy priority seems to perform the best among these three considered policies.

In the same panel, we also report the rate of ``correct'' decisions: accepting high-quality donors and rejecting low-quality ones. However, it may still be optimal for the patient to reject a high-quality organ, or accept a low-quality organ, as donor-patient compatibility is an important factor. We find that while the acceptance rate of high-quality donors remains consistently around 32\% across all three policies, the rejection rate for low-quality donors ranges from 42\% to 45\%, with the highest rejection rate under greedy priority.\footnote{Suppose all patients are homogeneous and base their decisions solely on their private signals. Our structural estimates suggest that the overall acceptance rate for high-quality donors is around 33\%, while the rejection rate for low-quality donors is approximately 52\%.} This is not surprising, as greedy priority aggregates the most negative information as patients move down the queue, increasing the likelihood of rejecting low-quality donors. Thus, for these patients to accept the offer, the utility of accepting must also be higher. This higher acceptance threshold could partially explain the improved acceptance utility under greedy priority.

Panel (C) of \autoref{tab:counter_efficiency} reports the average characteristics of patients who receive the organ. Compared to the current policy, greedy priority puts more weight on patient LAS and less on donor-patient distance when assigning priority. Specifically, the average LAS for patients under greedy priority is about 5\% higher than under the OPTN policy, while the average distance from the donor is about 56 nautical miles farther. Additionally, patients under greedy priority have the shortest waiting time among all three policies. In summary, greedy priority seems to benefit patients with high LAS, even if they are farther away from the donor or have waited a relatively short period, as they are deemed to be more likely to accept the offer.

\subsection{Alternative Information Treatments}
In this section, we focus on the OPTN priority policy but modify the information each patient observes when deciding whether to finally accept the offer. Our baseline is the current OPTN policy, which allows for social learning: each patient receives a private signal and observes the past rejections of higher priority patients. We then introduce two alternative informational treatments. The first treatment, referred to as \emph{no social learning}, completely eliminates the negative inference from social learning and aggregates no private information: each patient receives one private signal and acts as if they are the first in the priority queue. The second treatment, referred to as \emph{information sharing}, aggregates the maximum amount of private information: each patient receives their private signal and also observes the private signals of all higher-priority patients.

\begin{table}[t]
    \centering
    \caption{Counterfactual 2: Different Information Treatments Under the OPTN Priority policy}
    \begin{tabular}{lccc}
    \hline\hline
     &  (1) & (2) & (3)   \\ 
  Information  &  Social  & No Social & Information \\
   Treatment  &  Learning &  Learning &  Sharing \\
    \hline
    \multicolumn{4}{c}{\emph{Panel (A): Efficiency Outcomes}} \\
    Allocation Rate (\%) &  47.22	& 83.69 & 48.94

    \\
    Accepted Sequence Number & 6.91	& 7.33 & 2.49
   \\ \hline 
   \multicolumn{4}{c}{\emph{Panel (B): Welfare Outcomes}} \\
    Acceptance Utility  & 2.52	& 1.83 & 	2.71\\
    Total Acceptance Utility &  619.72	& 798.60 & 	692.09\\ 
   Acceptance Rate of High Quality (\%) & 32.63 & 	39.92	& 40.31 \\
   Rejection Rate of Low Quality (\%) & 44.53 & 	15.36	& 50.48
 \\
   \hline
\multicolumn{4}{c}{\emph{Panel (C): Accepted Patient Characteristics}} \\
 LAS & 59.12	& 52.63	 & 59.13
  \\
 Waiting Time (Month) & 5.35	& 7.02 & 6.49  \\
Primary blood type match & 1.61 & 1.57 & 1.61
  \\
Distance (NM) & 99.76 &94.82 & 	77.39
  \\ \hline
Provisional Yes & $\checkmark$  & $\checkmark$  & $\checkmark$  \\
Priority policy  & \multicolumn{3}{c}{OPTN Priority (Current)}   \\
    \hline\hline
    \end{tabular}
    \label{tab:counter_efficiency_info}
\end{table}

\autoref{tab:counter_efficiency_info} reports the outcomes of these informational treatments. Compared to the baseline where patients make negative inferences from past rejections, eliminating this negative inference nearly doubles the overall allocation rate, increasing it from 47\% to about 84\%. However, many of these additional acceptances result from ``incorrect'' decisions\textemdash accepting low-quality donors. In the no-social-learning treatment, the rejection rate of low-quality donors is only about 15\%, much lower than in treatments involving social learning or information sharing. In contrast, in the information-sharing treatment, where patients have access to the most information when making their decisions, the rejection rate of low-quality donors increases to about 51\%, and the resulting allocation rate is around 49\%, close to the 47\% observed with social learning. Thus, even though only negative information is aggregated, social learning achieves a similar allocation rate to that of full information transparency.

Furthermore, across all treatments, the average acceptance utility is lowest in the no-social-learning treatment, indicating poorer match quality due to the lack of information. Meanwhile, the average acceptance utility in the social learning treatment is similar to that in the information-sharing treatment. This suggests that social learning improves patient welfare by raising their threshold for accepting an offer. Additionally, having more information, whether negative or not, accelerates patients' decision-making. This is reflected in the average accepted sequence number: it is around 2.5 when patients share their private signals, increases to 6.9 when others' signals remain private but patients can infer from past rejections, and rises to 7.3 when there is no additional information beyond their own private signals.

We conduct two additional sets of counterfactual experiments on different information treatments, applying both greedy and reverse greedy priorities (see \autoref{tab:counter_efficiency_info_1} and \autoref{tab:counter_efficiency_info_2} in the appendix). The results there suggest that combining greedy priority with information sharing yields the best outcome overall. This combination not only maximizes patient welfare by achieving the highest average acceptance utility but also accelerates the allocation process by attaining the lowest average accepted sequence number. As seen before, one way to shift the current OPTN policy toward greedy priority is by placing greater emphasis on patient LAS and reducing the weight on donor-patient distance and waiting time when determining a patient's priority.

\section{Conclusion}
In this paper, we develop a structural model of sequential social learning, specifically adapted to the allocation of deceased donor lungs in the United States. Using data from a prominent organ procurement organization from 2019 to 2020, we estimate the model and find strong reduced-form evidence of social learning. Our counterfactual experiments demonstrate that social learning has a significant impact: under the current priority policy used by the OPTN, the average allocation rate nearly doubles when social learning is absent. However, this increase in organ allocation rate comes at the cost of lower patient utility per acceptance and a slower allocation process. Due to the heterogeneity in patients’ preferences for each organ\textemdash largely driven by donor-patient compatibility\textemdash the inefficiency caused by social learning remains relatively moderate. In fact, the allocation rate under social learning is only slightly lower than that achieved under information sharing, where patients have access to the most information.

When evaluating different priority policies, we find that greedy priority, which prioritizes patients with higher acceptance utility, delivers better outcomes in both patient welfare and allocation efficiency, even though social learning is most severe under this policy. When combined with policies that promote information sharing and transparency, greedy priority achieves the best overall outcome, effectively eliminating the negative effects of social learning.

\bibliographystyle{ecta}
\bibliography{reference,papers}

\begin{thebibliography}{18}
\newcommand{\enquote}[1]{``#1''}
\expandafter\ifx\csname natexlab\endcsname\relax\def\natexlab#1{#1}\fi

\bibitem[\protect\citeauthoryear{Agarwal, Ashlagi, Rees, Somaini, and Waldinger}{Agarwal et~al.}{2021}]{agarwal2021equilibrium}
\textsc{Agarwal, N., I.~Ashlagi, M.~A. Rees, P.~Somaini, and D.~Waldinger} (2021): \enquote{Equilibrium allocations under alternative waitlist designs: Evidence from deceased donor kidneys,} \emph{Econometrica}, 89, 37--76.

\bibitem[\protect\citeauthoryear{Agarwal, Ashlagi, Somaini, and Waldinger}{Agarwal et~al.}{2018}]{agarwal2018dynamic}
\textsc{Agarwal, N., I.~Ashlagi, P.~Somaini, and D.~Waldinger} (2018): \enquote{Dynamic incentives in wait list mechanisms,} in \emph{AEA Papers and Proceedings}, American Economic Association 2014 Broadway, Suite 305, Nashville, TN 37203, vol. 108, 341--347.

\bibitem[\protect\citeauthoryear{Banerjee}{Banerjee}{1992}]{banerjee1992simple}
\textsc{Banerjee, A.~V.} (1992): \enquote{A simple model of herd behavior,} \emph{The Quarterly Journal of Economics}, 107, 797--817.

\bibitem[\protect\citeauthoryear{Benvenuto and Aversa}{Benvenuto and Aversa}{2019}]{benvenuto2019should}
\textsc{Benvenuto, L.~J. and M.~Aversa} (2019): \enquote{Should We Accept This Offer? When Assessing Donor Lungs, Don’t Rely on Others,} \emph{Annals of the American Thoracic Society}, 16, 304--305.

\bibitem[\protect\citeauthoryear{Bikhchandani, Hirshleifer, and Welch}{Bikhchandani et~al.}{1992}]{BichHirshWelch:92}
\textsc{Bikhchandani, S., D.~Hirshleifer, and I.~Welch} (1992): \enquote{A theory of fads, fashion, custom, and cultural change as informational cascades,} \emph{Journal of Political Economy}, 992--1026.

\bibitem[\protect\citeauthoryear{Cipriani and Guarino}{Cipriani and Guarino}{2014}]{cipriani2014estimating}
\textsc{Cipriani, M. and A.~Guarino} (2014): \enquote{Estimating a structural model of herd behavior in financial markets,} \emph{American Economic Review}, 104, 224--51.

\bibitem[\protect\citeauthoryear{Cohen, Shults, Goldberg, Abt, Sawinski, and Reese}{Cohen et~al.}{2018{\natexlab{a}}}]{cohen2018position}
\textsc{Cohen, J., J.~Shults, D.~Goldberg, P.~Abt, D.~Sawinski, and P.~Reese} (2018{\natexlab{a}}): \enquote{Kidney allograft offers: Predictors of turndown and the impact of late organ acceptance on allograft survival,} \emph{American Journal of Transplantation}, 18, 391--401.

\bibitem[\protect\citeauthoryear{Cohen, Shults, Goldberg, Abt, Sawinski, and Reese}{Cohen et~al.}{2018{\natexlab{b}}}]{cohen2018kidney}
---\hspace{-.1pt}---\hspace{-.1pt}--- (2018{\natexlab{b}}): \enquote{Kidney transplant outcomes: position in the match-run does not seem to matter beyond other donor risk factors,} \emph{American Journal of Transplantation}, 18, 1577--1578.

\bibitem[\protect\citeauthoryear{De~Mel, Munshi, Reiche, Sabourian et~al.}{De~Mel et~al.}{2020}]{de2020herding}
\textsc{De~Mel, S., K.~Munshi, S.~Reiche, H.~Sabourian, et~al.} (2020): \enquote{Herding in Quality Assessment: An Application to Organ Transplantation,} Tech. rep., Faculty of Economics, University of Cambridge.

\bibitem[\protect\citeauthoryear{Gilroy, Cmunt, Kevin~Myer, Sonnenday, Preczewski, Stevens, Aguiar, and Swanson}{Gilroy et~al.}{2019}]{gilroy20192016}
\textsc{Gilroy, R., K.~Cmunt, M.~Kevin~Myer, C.~Sonnenday, L.~Preczewski, T.~Stevens, H.~Aguiar, and L.~Swanson} (2019): \enquote{Disrupting the Status Quo,} 2016 National Critical Issues Forum.

\bibitem[\protect\citeauthoryear{Goldberg, French, Lewis, Scott, Mamtani, Gilroy, Halpern, and Abt}{Goldberg et~al.}{2016}]{goldberg2016liver}
\textsc{Goldberg, D.~S., B.~French, J.~D. Lewis, F.~I. Scott, R.~Mamtani, R.~Gilroy, S.~D. Halpern, and P.~L. Abt} (2016): \enquote{Liver transplant center variability in accepting organ offers and its impact on patient survival,} \emph{Journal of hepatology}, 64, 843--851.

\bibitem[\protect\citeauthoryear{Hackmann, English, and Kizer}{Hackmann et~al.}{2022}]{hackmann2022improving}
\textsc{Hackmann, M., R.~A. English, and K.~W. Kizer} (2022): \enquote{Improving Procurement, Acceptance, and Use of Deceased Donor Organs,} in \emph{Realizing the Promise of Equity in the Organ Transplantation System}, National Academies Press (US).

\bibitem[\protect\citeauthoryear{Harhay, Porcher, Thabut, Crowther, DiSanto, Rubin, Penfil, Bing, Christie, Diamond et~al.}{Harhay et~al.}{2019}]{harhay2019donor}
\textsc{Harhay, M.~O., R.~Porcher, G.~Thabut, M.~J. Crowther, T.~DiSanto, S.~Rubin, Z.~Penfil, Z.~Bing, J.~D. Christie, J.~M. Diamond, et~al.} (2019): \enquote{Donor lung sequence number and survival after lung transplantation in the United States,} \emph{Annals of the American Thoracic Society}, 16, 313--320.

\bibitem[\protect\citeauthoryear{Mu{\~n}oz-Rodriguez}{Mu{\~n}oz-Rodriguez}{2021}]{munoz2021incentive}
\textsc{Mu{\~n}oz-Rodriguez, E.} (2021): \enquote{Incentive-Compatible Triaging in Deceased Donor Transplantation,} \emph{Available at SSRN 3805693}.

\bibitem[\protect\citeauthoryear{Newham and Midjord}{Newham and Midjord}{2018}]{newham2018herd}
\textsc{Newham, M. and R.~Midjord} (2018): \enquote{Herd behavior in FDA committees: A structural approach,} DIW Berlin Discussion Paper No. 1744.

\bibitem[\protect\citeauthoryear{OPTN}{OPTN}{2019}]{OPTN_2019}
\textsc{OPTN} (2019): \enquote{Organ Procurement and Transplantation Network Policies,} Tech. rep., OPTN.

\bibitem[\protect\citeauthoryear{Shi and Yin}{Shi and Yin}{2022}]{shi2022eliminating}
\textsc{Shi, P. and J.~Yin} (2022): \enquote{Eliminating Waste in Cadaveric Organ Allocation,} \emph{USC Marshall School of Business Research Paper Sponsored by iORB}.

\bibitem[\protect\citeauthoryear{Zhang}{Zhang}{2010}]{zhang2010sound}
\textsc{Zhang, J.} (2010): \enquote{The sound of silence: Observational learning in the US kidney market,} \emph{Marketing Science}, 29, 315--335.

\end{thebibliography}

\appendix

\section{Appendix}
\numberwithin{equation}{section}
\subsection{Additional Tables and Figures}
\begin{table}[h!] 
    \centering \caption{Estimation Results}
    \begin{tabular}{lcc}
    \hline \hline
      Parameter   & Estimate  & S.D. \\
      \hline
     $\mu$      & 0.958 & --\\
      $\alpha$   & 0.850 & 0.039 \\
      $p$   & 0.383 & 0.046 \\
      $\gamma$  & 4.934 & 0.005 \\
      $\boldsymbol{\tilde{\beta}}$  & & \\
   \quad \emph{Patient Characteristics} & & \\
\quad   \quad LAS & 0.033 & 0.001 \\
  \quad  \quad Waiting time (month) &  -0.024& 0.002 \\
  \quad   \quad BMI & 0.145
 & 0.001 \\
  \quad  \quad Female   & -0.053
 & 0.005 \\
 \quad \quad Diabetic  & 0.130
 & 0.244 \\
 \quad \quad Previous lung transplant & 0.439 & 0.413 \\
 \quad \emph{Donor Characteristics} & &  \\
 \quad \quad P/F ratio & 0.001& 0.000 \\
 \quad \quad Age (year) & 0.005 & 0.001 \\
 \quad \quad Weight (kg) & -0.070& 0.001 \\
 \quad \quad Height (cm) & 0.071& 0.001 \\
 \quad \quad History of IV drug use & -0.014& 0.344 \\
 \quad \quad History of heavy alcohol consumption & -0.034 & 0.040 \\
 \quad  \quad Increasing risk status & -0.149& 0.111\\
 \quad   \emph{Donor-Patient Pair Characteristics} & & \\
  \quad    \quad Primary blood type match & 0.775 & 0.006
 \\  
  \quad   \quad Distance (per NM) & -0.003 & 0.001\\
  \quad   \quad Age difference & 0.027 &  0.001\\
  \quad   \quad Height difference & 0.076 & 0.002\\
   \quad   \quad Weight difference & -0.064 & 0.001\\
         \hline \hline \\ [-3ex] 
          \end{tabular}
    \label{tab:estimation_full} 
  \begin{minipage}{11cm} ~\\
    \small {\emph Note: This table shows the full set of estimates for our structural model. The standard deviations are obtained through the BHHH estimator.}  
  \end{minipage}
\end{table}

\begin{table}[ht]
    \centering
    \caption{Counterfactual 3: Different Information Treatments Under the Greedy Priority Policy}
    \begin{tabular}{lccc}
    \hline\hline
     &  (1) & (2) & (3)   \\ 
  Information  &  Social  & No Social & Information \\
   Treatment  &  Learning &  Learning &  Sharing \\
    \hline
    \multicolumn{4}{c}{\emph{Panel (A): Efficiency Outcomes}} \\
    Allocation Rate (\%) &  46.26	& 83.69 & 50.10

    \\
    Accepted Sequence Number & 5.13	& 6.46 & 2.34
   \\ \hline 
   \multicolumn{4}{c}{\emph{Panel (B): Welfare Outcomes}} \\
    Acceptance Utility  & 2.75	& 1.99 & 	2.95\\
    Total Acceptance Utility &  662.27	& 867.18 & 	770.70\\ 
   Acceptance Rate of High Quality (\%) & 32.44 & 	39.73	& 40.31 \\
   Rejection Rate of Low Quality (\%) & 45.30 & 	15.16	& 49.33
 \\
   \hline
\multicolumn{4}{c}{\emph{Panel (C): Accepted Patient Characteristics}} \\
 LAS & 64.24	& 57.52	 & 64.24
  \\
 Waiting Time (Month) & 3.03	& 3.97 & 2.93  \\
Primary blood type match & 1.60 & 1.58 & 1.62
  \\
Distance (NM) & 155.73 &150.50 & 	144.66
  \\ \hline
Provisional Yes & $\checkmark$  & $\checkmark$  & $\checkmark$  \\
Priority policy  & \multicolumn{3}{c}{Greedy Priority}   \\
    \hline\hline
    \end{tabular}
    \label{tab:counter_efficiency_info_1}
\end{table}

\begin{table}[ht]
    \centering
\caption{Counterfactual 4: Different Information Treatments Under the Reverse Greedy Priority Policy}
    \begin{tabular}{lccc}
    \hline\hline
     &  (1) & (2) & (3)   \\ 
  Information  &  Social  & No Social & Information \\
   Treatment  &  Learning &  Learning &  Sharing \\
    \hline
    \multicolumn{4}{c}{\emph{Panel (A): Efficiency Outcomes}} \\
    Allocation Rate (\%) &  49.33	& 83.30 & 45.11

    \\
    Accepted Sequence Number & 15.49	& 17.88 & 6.79
   \\ \hline 
   \multicolumn{4}{c}{\emph{Panel (B): Welfare Outcomes}} \\
    Acceptance Utility  & 1.48	& 1.17 & 	1.49\\
    Total Acceptance Utility &  381.03	& 507.89 & 	350.06\\ 
   Acceptance Rate of High Quality (\%) & 32.63 & 	39.54	& 40.31 \\
   Rejection Rate of Low Quality (\%) & 42.42 & 	15.36	& 54.32
 \\
   \hline
\multicolumn{4}{c}{\emph{Panel (C): Accepted Patient Characteristics}} \\
 LAS & 38.15	& 38.51	 & 37.43
  \\
 Waiting Time (Month) & 8.27	& 8.80 & 10.02  \\
Primary blood type match & 1.54 & 1.52 & 1.54
  \\
Distance (NM) & 339.75 &300.93 & 445.80
  \\ \hline
Provisional Yes & $\checkmark$  & $\checkmark$  & $\checkmark$  \\
Priority policy  & \multicolumn{3}{c}{Reverse Greedy Priority}   \\
    \hline\hline
    \end{tabular}  \label{tab:counter_efficiency_info_2}
\end{table}

\end{document}